\documentclass[conference,letterpaper]{IEEEtran}

\usepackage[utf8]{inputenc} 
\usepackage[T1]{fontenc}
\usepackage{url}
\usepackage{ifthen}
\usepackage{cite}
\usepackage[cmex10]{amsmath} 
\usepackage{url}
\usepackage{lettrine}
\usepackage{framed}
\usepackage{color}
\usepackage{amssymb}
\usepackage{algorithm}
\usepackage{algpseudocode}
\usepackage{dsfont}
\usepackage{booktabs} 
\usepackage{nicefrac}
\usepackage{framed}
\usepackage{bm}
\usepackage{bbm}
\usepackage{ifthen}
\usepackage{cite}
\usepackage{graphicx}
\usepackage{dsfont}
\usepackage{framed}
\usepackage{bm}
\usepackage{overpic}
\usepackage{xcolor}
\usepackage{subcaption}
\usepackage{hyperref}
\newtheorem{theorem}{Theorem}

\newtheorem{lemma}{Lemma}
\usepackage{xr}
\newcommand{\cmt}[1]{\textcolor{red}{\textbf{(AS: #1)}}}

\begin{document}
\title{Minimizing Queue Length Regret for Arbitrarily Varying Channels} 


\author{%
  \IEEEauthorblockN{G Krishnakumar}
    \IEEEauthorblockA{Dept. of Electrical Engineering \\ Indian Institute of Technology Madras\\
                    Chennai 600036, India\\ 
                    Email: \textrm{krishnakumar97@smail.iitm.ac.in}
                    }
                    \and 
    \IEEEauthorblockN{Abhishek Sinha}                
  \IEEEauthorblockA{School of Technology and Computer Science \\
                    Tata Institute of Fundamental Research\\
                    Mumbai 400005, India\\
                    Email: \textrm{abhishek.sinha@tifr.res.in}}}

\maketitle

\begin{abstract}
   We consider an online channel scheduling problem for a single transmitter-receiver pair equipped with $N$ arbitrarily varying wireless channels. The transmission rates of the channels might be non-stationary and could be controlled by an oblivious adversary. At every slot, incoming data arrives at an infinite-capacity data queue located at the transmitter. A scheduler, which is oblivious to the current channel rates, selects one of the $N$ channels for transmission. At the end of the slot, the scheduler only gets to know the transmission rate of the selected channel. The objective is to minimize the \emph{queue length regret}, defined as the difference between the queue length at some time $T$ achieved by an online policy and the queue length obtained by always transmitting over the single best channel in hindsight. We propose a \emph{weakly adaptive} Multi-Armed Bandit (MAB) algorithm for minimizing the queue length regret in this setup. 
   Unlike previous works, we do not make any stability assumptions about the queue or the arrival process. Hence, our result holds even when the queueing process is unstable. Our main observation is that the queue length regret can be upper bounded by the regret of a MAB policy that competes against the best channel in hindsight uniformly over \emph{all} sub-intervals of $[T]$. As a technical contribution of independent interest, we then propose a weakly adaptive adversarial MAB policy which achieves $\tilde{O}(\sqrt{N}T^{\nicefrac{3}{4}})$ regret with high probability, implying the same bound for queue length regret. 
 
   
   \end{abstract}
\section{Introduction} \label{intro}
Dynamic channel selection in wireless systems with limited feedback has been extensively studied in the literature \cite{takeuchi2020dynamic, combes2014dynamic}. 
In this problem, there is a set of $N$ available channels whose quality may fluctuate from time to time due to wireless fading, interference, adversarial jamming, or other impairments.
The problem is to simultaneously estimate the channel qualities (exploration) and transmit over the best channel in hindsight (exploitation). Several prior studies employed the standard non-adaptive Multi-Armed Bandit (MAB) framework to learn the best channel in hindsight by minimizing the regret of cumulative transmission rates, which corresponds to maximizing the transmission opportunities \cite{ch-selection-ref1, ref2, survey-MAB}. However, often in practice, the primary goal is to minimize QoS metrics such as delay or queue lengths (which are equivalent whenever Little's law holds \cite{bertsekas2021data}). Note that queue length evolves through a non-linear recursive dynamics and becomes independent of the transmission rates whenever the queue becomes empty. Hence, a direct application of non-adaptive MAB policies does not optimize for the objective of interest. Indeed, the paper \cite{stahlbuhk2021learning} showed that the standard MAB-based policies have a sub-optimal performance in a stochastic setting. Furthermore, a large number of works in the wireless scheduling literature assume a stationary channel process and data arrival model for analytical tractability \cite{krishnasamy2016regret, stahlbuhk2021learning, hsu2007modeling}. However, these simple models are not adequate for many practical scenarios featuring non-stationary channels, \emph{e.g.,} 5G mmWave \cite{sinha2022optimizing, wu2017general}, or bursty arrival process, \emph{e.g.,} internet traffic \cite{nain2002impact}. 
  
\begin{figure}
	\includegraphics[scale=0.5]{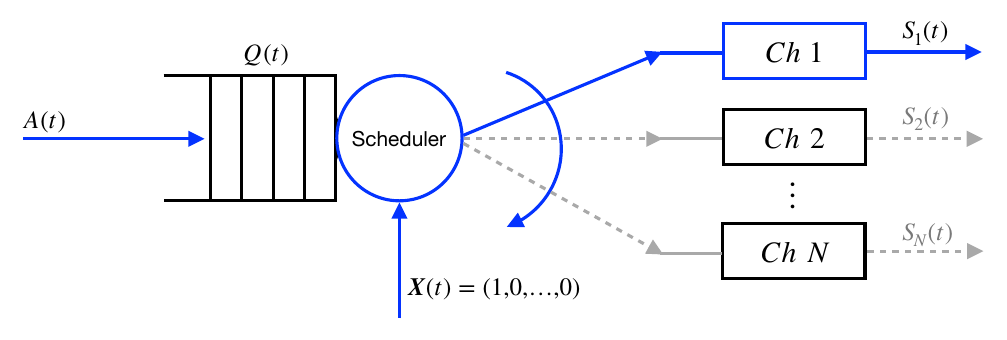}
	\caption{\small{Schematic of the scheduling problem with $N$ arbitrary varying channels.}}
	\label{sched-fig}
\end{figure}

In this paper, we attempt to address both of the above issues by considering an adversarial data arrival and channel model. Our objective is to minimize the maximum queue length within a horizon of  $T$ slots. As mentioned above, due to the non-separable nature of the queueing process, minimizing the queue length regret is substantially more challenging than just running a non-adaptive MAB algorithm.
To solve this problem, in this paper, we make the following contributions:

\begin{enumerate}
\item We reduce the problem of minimizing the queue length regret to designing a weakly adaptive MAB policy which controls the regret over \emph{all} sub-intervals. This reduction holds without any assumption on the arrival process, service process, or the stability status of the queue.
	\item 
We then propose a weakly adaptive MAB policy which achieves $\tilde{O}(T^{\nicefrac{3}{4}})$ regret for all sub-intervals of $[T]$ with high probability (Theorem \ref{ogd-mab}). This result might be of independent interest.
\item Finally, we demonstrate that our policy outperforms two previously proposed scheduling policies in numerical simulations.

 \end{enumerate}
 	
\section{Problem Formulation} \label{analysis}
We consider a single transmitter-receiver pair which can communicate over any of the $N$ given wireless channels. See Figure \ref{sched-fig} for a schematic diagram of the system.
Let $Q(t)$ denote the length of the data queue at slot $t.$
The time evolution of $Q(t)$ follows the standard Lindley recursion:
\begin{eqnarray} \label{q-ev}
	Q(t)= \big(Q(t-1)+A(t)-\langle \bm{S}(t), \bm{X}(t) \rangle \big)^+, ~Q(0)=0,
\end{eqnarray}
where $A(t) \geq 0$ is the amount of data that arrives on slot $t$, $\bm{S}(t)$ is an $N$-dimensional non-negative service rate vector denoting the instantaneous transmission rates of the channels, and $\bm{X}(t)$ is an $N$-dimensional unit vector, which denotes the channel used by the policy for transmission on slot $t$. Note that the online scheduling policy must make the scheduling decision $\bm{X}(t)$ without the knowledge of the current service rates $\bm{S}(t)$ and any future arrival or transmission rate information. We will see that our proposed policy is also oblivious to the past arrivals and the queue length process $\{Q(t)\}_{t \geq 1}$. 
\paragraph*{Assumption}
We make the standard scaling assumption that the magnitude of each component of the transmission rate vector is bounded by one, \emph{i.e.,} $||\bm{S}(t)||_\infty \leq 1, \forall t \geq 1.$ We also assume that $\{\bm{S}(t)\}_{t \geq 1}$ is controlled by an \emph{oblivious} adversary, \emph{i.e.,} its values are chosen at the beginning ($t=0$).

Notably, we do not impose \emph{any} restriction on the arrival process $\{A(t)\}_{t\geq 1},$ which could take arbitrarily large values. We also do not make any assumption about the relative magnitudes of the arrival and service processes and, hence, the stability of the queue. This is in sharp contrast with the previous works on this problem, which assume the stability of the queue  \cite{stahlbuhk2021learning, krishnasamy2016regret}, and the heavy-traffic model \cite{krishnasamy2016regret}. 
\paragraph*{Feedback model} If at the $t$\textsuperscript{th} slot, the scheduling policy selects the $i$\textsuperscript{th} channel, then only the $i$\textsuperscript{th} component of the transmission rate vector $S_i(t)$ is revealed to the policy at the end of the slot. 
Following \cite{stahlbuhk2021learning}, we assume that if the queue is empty, the policy can send a dummy packet to obtain the current transmission rate information on the selected channel. 
\subsection*{Performance metric: Queue Length Regret}
Let $\{Q^i(t)\}_{t\geq 1}$ be the queue length process under a fixed policy which always schedules the $i$\textsuperscript{th} channel. In other words, its action is given by $\bm{X}^i(t)=\bm{e}_i, \forall t \geq 1,$ where $\bm{e}_i$ is the $N$-dimensional unit vector whose $i$\textsuperscript{th} component is $1$ and the rest of the components are zero. Then the worst-case \emph{Queue Length Regret} up to time $T$ under a policy $\pi$ is defined as: 
\begin{eqnarray} \label{regret-def}
	R_Q^\pi(T) := \max_{t \in [T]} \big(\max_{i \in [N]} \big(Q^\pi(t) - Q^i(t)\big)\big).
\end{eqnarray} 
We use the superscript $\pi$ to emphasize that the channel allocation vectors $\{\bm{X}(t)\}_{t\geq 1}$ are controlled by the scheduling policy $\pi.$  
In general, $R^\pi(T)$ could be a random variable owing to the randomness of the policy. The regret measure \eqref{regret-def} has been used earlier for studying the performance of scheduling policies under stochastic assumptions \cite{krishnasamy2016regret}. Readers should distinguish this measure from the cumulative queue length regret metric studied in \cite{stahlbuhk2021learning}.

Definition \eqref{regret-def} is comparable to the standard definition of regret in the online learning literature \cite{cesa2006prediction} with an additional complication arising from the fact that the queue lengths, being a stateful system, \emph{cannot} be expressed as a sum of cost components across time. Furthermore, since we are considering the worst-case queue length regret within a horizon, the metric \eqref{regret-def} is stronger than the terminal queue length metric considered in \cite{krishnasamy2016regret}.
 	
\begin{table*}[t]
\label{comp-table}
\begin{center}
\begin{small}
\begin{sc}
\begin{tabular}{lcccccr}
\toprule
Reference &  Performance Metric & Assumptions& Regret  & Policy & \\
\midrule
\cite{stahlbuhk2021learning} &  $\sum_{t=1}^T Q(t)$& IID Arrivals, Stability$^\star$& $O(1)$& $Q(t)$- dependent  & \\\cite{krishnasamy2016regret}& $Q(T)$& IID Arrivals, Stability$^\star$, Heavy load& $\tilde{O}(T^{-1})$&$Q(t)$- independent & \\
\textbf{This work}&  $\max_{t\leq T} Q(t)$&--- & $\tilde{O}(T^{\nicefrac{3}{4}})$ &  $Q(t)$-independent & \\
\bottomrule
\end{tabular}
\end{sc}
\end{small}
\end{center}
\caption{\small{Summary of the results on minimizing the queue length regret. $^\star$Stability refers to the assumption that there exists a policy which stabilizes the queue.}}
\end{table*}

\section{An Online Scheduling Algorithm} \label{q-reg-analysis}
Our main observation is that the queue length regret, given by Eqn.\ \eqref{regret-def}, can be upper bounded by the regret of a weakly adaptive adversarial MAB algorithm, which competes with the best arm in hindsight across all sub-intervals. To obtain this result, we first expand the Lindley recursion \eqref{q-ev} to obtain the following standard representation of queue length \cite[Lemma 1.1]{ganesh2004big}:
\begin{eqnarray} \label{q-lindley}
	Q(t) = \max_{1\leq t' \leq t} \sum_{\tau=t'}^t \big(A(\tau) -\langle \bm{S}(\tau), \bm{X}(\tau) \rangle\big), ~t\geq 1.
\end{eqnarray}
For completeness, we provide a proof of this result in Lemma \ref{lindley_proof} in the Appendix.

Now, we fix a scheduling policy $\pi$ and a sample path. For a given horizon length $T,$ let $t \in [T]$ achieve the outer $\arg\max$ in Eqn.\ \eqref{regret-def} and let $i^\star_t$ be the corresponding best channel in hindsight, achieving the inner $\arg\max$ in \eqref{regret-def}. Furthermore, for this $t,$ let $t' \leq t$ achieve the $\arg \max$ in Eqn.\ \eqref{q-lindley} for $Q^\pi(t)$.
We can now upper bound the queue length regret \eqref{regret-def} as follows: 
\begin{align} \label{lindley}
	&R_Q^\pi(T) \nonumber \\
	&=  Q^\pi(t) - Q^{i_t^\star}(t) \nonumber \\
	&\leq  \sum_{\tau=t'}^t \big(A(\tau) -\langle \bm{S}(\tau), \bm{X}^\pi(\tau) \rangle\big) - \sum_{\tau=t'}^t \big(A(\tau) \nonumber \\
	&-\langle \bm{S}(\tau), \bm{X}^{i_t^\star}(\tau) \rangle\big) \nonumber \\
	&=\sum_{\tau=t'}^t \langle \bm{S}(\tau), \bm{X}^{i_t^\star}(\tau) \rangle - \langle \bm{S}(\tau), \bm{X}^{\pi}(\tau) \rangle \nonumber \\
	&\leq \max_{ [t',t] \subseteq [T]} \max_{i^\star \in [N]} \bigg(\sum_{\tau=t'}^t \langle \bm{S}(\tau), \bm{X}^{i^\star}(\tau) - \bm{X}^{\pi}(\tau) \rangle\bigg).  
\end{align}
Interestingly, the bound \eqref{lindley} does not involve the arrival process $\{A(t)\}_{t \geq 1}$ at all. As a result, unlike previous works \cite{stahlbuhk2021learning, krishnasamy2016regret}, we do not need to make any assumption about the arrival process and, consequently, about the stability of the queue. 

Motivated by the above observation, our task now is to design an online scheduling policy $\pi$ which learns the best channel in hindsight by minimizing the RHS in \eqref{lindley}. This is equivalent to an adversarial Multi-armed Bandit (MAB) problem where the transmission rates $\bm{S}(t)$ of $N$ different channels are identified with the gains $\bm{g}(t)$ of $N$ different arms of the bandit (See Section \ref{prelim}). However, since \eqref{lindley} involves a maximum over all subintervals, the proposed MAB policy must achieve a \emph{weakly adaptive} regret bound by uniformly controlling the regret over \emph{all subintervals} of $[T]$ w.h.p.  In other words, the desired online learner must be stronger than a standard non-adaptive regret minimizer (\emph{e.g.,} EXP3), which only controls the terminal regret. The following result formally establishes the hardness of the problem.

\paragraph*{Lower Bound} We now show that the queue length regret is lower bounded by the regret lower bound for the standard adversarial MAB problem, which is well-known to be $\Omega(\sqrt{NT})$ w.h.p. \cite{lattimore2020bandit}. To see this, consider an arrival process such that the number of arrivals on each slot is at least $1,$ \emph{i.e.,} $A(t) \geq 1, \forall t\geq 1.$ Since under our assumption $||\bm{S}(t)||_\infty \leq 1, \forall t\geq 1,$ for any policy we have $A(t) \geq \langle \bm{S}(t), \bm{X}(t)\rangle, \forall t\geq 1.$ Hence, with this particular choice of the arrival process, the $\max(0,\cdot)$ operator in the Lindley recursion \eqref{q-ev} becomes superfluous and the queueing dynamics becomes linear. As a result, after the cancellation of the common terms, the queue length regret \eqref{regret-def} becomes exactly equal to the regret for cumulative transmission rates, and hence, the lower bound follows.  

Motivated by the above, in the following section, we propose a weakly adaptive MAB policy which achieves $\tilde{O}(\sqrt{N}T^{\nicefrac{3}{4}})$ regret over all sub-intervals of $[T]$ w.h.p.



\section{A Weakly Adaptive Algorithm for Adversarial Bandits} \label{prelim}

Adversarial Multi-armed Bandits (MAB) is a widely studied framework for decision-making under uncertainty with limited feedback. We refer the reader to standard references \cite{lattimore2020bandit, bubeck2012regret} for details on the MAB framework and related algorithms. To set up the notations, we briefly recall that MAB is a repeated game played with $N$ arms between an online algorithm and an adversary for $T$ rounds. At the beginning of the game, an oblivious adversary secretly selects a sequence of non-negative gain vectors $\{\bm{g}_t\}_{t\geq 1}$ for all $T$ rounds such that $||\bm{g}_t||_2 \leq G, \forall t\geq 1$. On every round $t$ an online policy $\pi$ pulls an arm 
$J^\pi_t \in [N]$ and receives a reward of $g_{t, J^\pi_t}$ for that round. For notational convenience, we use $\bm{J}^\pi_t$ to denote the one-hot encoded $N$-dimensional vector corresponding to the categorical variable $J^\pi_t$. The \emph{regret} of the policy $\pi$ for a  sub-interval $I \subseteq [T]$ is defined as: 
\begin{eqnarray} \label{reg-def}
	R^\pi_I= \max_{\bm{J}^\star} \sum_{t \in I}\langle \bm{g}_t,\bm{J}^\star \rangle - \sum_{t \in I} \langle \bm{g}_t, \bm{J}^\pi_t \rangle,   
\end{eqnarray}
where $\bm{J}^\star$ is the best constant benchmark action which always selects one particular arm throughout the sub-interval $I$. The standard objective is to minimize the end-to-end cumulative regret $R^\pi_{[T]}.$ Note that the regret defined in \eqref{reg-def} for MAB is a random quantity, and, in this paper, we will be concerned with regret bounds that hold with high probability.

A policy $\pi$ is called \emph{weakly adaptive} if $\max_{I \subseteq [T]} R^\pi_I$ is sublinear in $T$ w.h.p. where the maximum is taken over all sub-intervals. This must be contrasted with the so-called \emph{strongly adaptive} policies where $\max_{I \subseteq [T]} R^\pi_I $ is required to be sublinear in the length of the interval $|I|.$ The paper \cite{daniely2015strongly} showed that there cannot exist any strongly adaptive online learning algorithm in the bandit feedback setting. The problem of the existence of a weakly adaptive bandit algorithm was left as an open problem. In this paper, we design a weakly adaptive MAB algorithm and show that it suffices for our problem.

We first describe the ubiquitous Online Gradient Descent (OGD) policy specialized to the \emph{full-information} version of the MAB problem, known as the \emph{expert problem} in the literature \cite{cesa2006prediction}. This will be used in our proposed MAB algorithm.   
\begin{algorithm}
\caption{\textsc{Online Gradient Descent for the Expert Problem}}
\label{ogd.f}

\begin{algorithmic}[1]
\State \textbf{Parameters:} Step size $\eta = \frac{\sqrt{2}}{G \sqrt{T}},$ $(N-1)$-dimensional Probability simplex $\Delta_N,$ Euclidean projection operator $\textsc{Proj}_{\Delta_N}(\cdot)$ onto the simplex $\Delta_N.$
\State \textbf{Initialization:} Initialize $p_1 \in \Delta_N$ arbitrarily
\For{$t = 1, 2, \dots, T$}
    \State Choose an arm according to the distribution $\bm{p}_t,$ observe the full gain vector $\bm{g}_t$.
        \State Update the probability distribution over arms:
    \[
    \bm{p}_{t+1} = \textsc{Proj}_{\Delta_N}\big(\bm{p}_t + \eta \bm{g}_t \big). ~~~~(\star)
    \]
 
\EndFor
\end{algorithmic}
\end{algorithm}
\paragraph*{Remark (Computational Complexity)} The most computationally expensive operation Algorithm \ref{ogd.f} is the projection step $(\star)$ which is performed every round. However, it is a well-known fact that the Euclidean projection on the simplex $\Delta_N$ can be efficiently computed in time linear in $N$ \cite{chen2011projection}.

The following result shows that OGD already enjoys a weakly adaptive $O(\sqrt{T})$ regret bound.
\begin{theorem} \label{ogd-w}
	The Online Gradient Descent (OGD) policy for the expert problem in the full information setting, described in Algorithm \ref{ogd.f} is weakly adaptive and achieves the following bound for every sub-interval $I \subseteq [T]$:
	\[\max_{I \subseteq [T]} R^\textrm{OGD}_I = G\sqrt{2T}. \]
\end{theorem}
 
\begin{IEEEproof}
	Fix any sub-interval $I \subseteq [T]$ and the corresponding (one-hot encoded) fixed best arm in hindsight $\bm{p}^\star_I \in \Delta_N.$ Note that the benchmark $\bm{p}^\star_I$ could depend on the sub-interval $I.$ We have 
	\begin{eqnarray*}
		&&||\bm{p}_{t+1} - \bm{p}_I^\star||^2 \\
		&=& ||\textsc{Proj}_{\Delta_N} (\bm{p}_t+ \eta \bm{g}_t) - \bm{p}^\star_I||^2 \\
		&\stackrel{(a)}{\leq} & ||\bm{p}_t - \bm{p}^\star_I + \eta \bm{g}_t ||^2 \\
		&=& ||\bm{p}_t - \bm{p}^\star_I||^2 + 2 \eta \langle \bm{p}_t - \bm{p}^\star_I, \bm{g}_t \rangle  + \eta^2 ||\bm{g}_t||^2 
	\end{eqnarray*}
	where the inequality (a) follows from the Pythagorean theorem of Euclidean projection \cite{bertsekas}. Thus, the instantaneous regret on the $t$\textsuperscript{th} round can be upper bounded as 
	\begin{eqnarray*}
		\langle \bm{p}^\star_I - \bm{p}_t, \bm{g}_t \rangle \leq \frac{||\bm{p}_t- \bm{p}^\star_I ||^2- ||\bm{p}_{t+1}- \bm{p}^\star_I||^2}{2\eta}+\frac{\eta}{2}G^2,   
	\end{eqnarray*} 
	where we have used the fact that $||\bm{g}_t|| \leq G.$ Summing up the above inequalities for all $t\in I,$ we obtain 
\begin{eqnarray*}
	R_I^{\textrm{OGD}} \leq \frac{||\bm{p}_{I_1}- \bm{p}^\star_I||^2}{2\eta } + \frac{\eta}{2}G^2|I| \stackrel{(a)}{\leq}   \frac{1}{\eta} + \frac{\eta G^2 T}{2},
\end{eqnarray*}	
where in (a) we have used the fact that for any two probability distributions $\bm{p}$ and $\bm{q}$ on the simplex $\Delta_N,$ $||\bm{p}-\bm{q}||^2 = \sum_{i} (p_i-q_i)^2 \leq \sum_i |p_i-q_i| \leq \sum_i p_i + \sum_i q_i = 2$ and $|I|\leq T.$ Substituting for $\eta= \frac{\sqrt{2}}{G\sqrt{T}},$ the above bound reduces to 
\[	R_I^{\textrm{OGD}} \leq G\sqrt{2T}, ~~ \forall I \subseteq [T] . \]  
The final result follows upon taking maximum over all sub-intervals $I \subseteq [T].$
\end{IEEEproof}
Our main result shows that Algorithm \ref{mab}, which uses the full-information Algorithm \ref{ogd.f} as a subroutine along with random exploration, achieves an $\tilde{O}(T^{\nicefrac{3}{4}})$ weakly adaptive regret bound with high probability in the bandit feedback setting.

\begin{theorem} \label{ogd-mab}
	Algorithm \ref{mab} is a weakly adaptive MAB policy and achieves the following regret bound uniformly over every sub-interval $I \subseteq [T]$, where $T \geq N^2$
	\begin{eqnarray*}
		 	\max_{I \subseteq [T]} R_I^{\textrm{ALG2}} \leq 3 \sqrt{N}T^{\nicefrac{3}{4}} \big(1+ \sqrt{\ln \frac{3NT^2}{\delta}}\big),  \textrm{w.p. at least } 1-\delta.
		 \end{eqnarray*}
		 \end{theorem} 
\paragraph*{Corollary} Discussion in Section \ref{q-reg-analysis} implies that Algorithm \ref{mab} achieves the above bound for the queue length regret metric \eqref{regret-def}.

\begin{algorithm}
\caption{\textsc{A Weakly Adaptive MAB algorithm}}
\label{mab}

\begin{algorithmic}[1]
\State \textbf{Parameters:} Exploration probability \[\gamma = \min (1, \sqrt{N}T^{-\nicefrac{1}{4}})\]


\For{$t = 1, 2, \dots, T$}
    \State Pull arm $J_t$ independently at random as follows:
    \begin{eqnarray*}
    	J_t \sim \begin{cases}
    		\textrm{Unif}[N], ~~ \textrm{w.p.} ~\gamma \\
    		\bm{p}_t, ~\textrm{where}~  \bm{p}_t =\textrm{OGD}(\bm{\hat{g}}_{t-1}, \eta=\frac{\sqrt{2}\gamma}{N\sqrt{T}}), \textrm{w.p.~} 1-\gamma  
    	\end{cases}
    \end{eqnarray*} 
    \State Observe $g_{t, J_t}$
    \State Define the estimated gain vector for round $t$ as
    \begin{eqnarray} \label{est}
    	\hat{g}_{t,i} = \frac{g_{t,i}}{\mathbb{P}(J_t=i)} \mathds{1}(J_t=i), ~ i \in [N].
    \end{eqnarray}  
    where $\mathbb{P}(J_t=i)= (1-\gamma) p_{t,i}+\gamma/N.$
    \EndFor
\end{algorithmic}
\end{algorithm}

\begin{IEEEproof}
 Fix any sub-interval $I \subseteq [T],$ an arbitrary benchmark $\bm{p}^\star \in \Delta_N,$ and a confidence parameter $\delta' >0$.  
Since $\mathbb{P}(J_t =i ) \geq \frac{\gamma}{N}$ and $g_{t,i} \leq 1,$ from Eqn. \eqref{est}, we have $\hat{g}_{t,i} \leq \frac{N}{\gamma}, \forall t,i.$ Furthermore, since only one component of the estimated gain vector $\hat{\bm{g}}_t$ is non-zero, we have  $||\bm{\hat{g}}_t||_2 \leq \frac{N}{\gamma}, \forall t.$ Hence, using Theorem \ref{ogd-w}, we have the following bound, which holds for every sample path:
\begin{eqnarray} \label{ogd-bd}
	\sum_{t \in I} \langle \hat{\bm{g}}_t, \bm{p}^\star \rangle - \sum_{t \in I}\langle \hat{\bm{g}}_t, \bm{p}_t \rangle \leq  \frac{N}{\gamma}\sqrt{2T}.
\end{eqnarray} 
Let $\{\mathcal{F}_\tau\}_{\tau\geq 1}$ be the associated natural filtration. It can be easily verified that the gain estimator \eqref{est} is conditionally unbiased as $\mathbb{E}(\hat{g}_{t,i}|\mathcal{F}_{t-1})= g_{t,i}, \forall i,t.$ 
Furthermore, since the sequence $\{\bm{p}_\tau\}_{\tau \geq 1}$ is previsible, \emph{i.e.,} $\bm{p}_t \in \mathcal{F}_{t-1}, \forall t,$ it follows that the sum $\sum_{\tau}\langle \hat{\bm{g}}_\tau - \bm{g}_\tau, \bm{p}_\tau \rangle$ is a zero-mean Martingale difference sequence. 
Hence, using Azuma's inequality \cite[Theorem 6.3.3]{ross1995stochastic},
we conclude that with probability at least $1-\delta',$ we have: 
\begin{eqnarray} \label{mg-bd1}
	\sum_{t \in I} \langle \bm{\hat{g}}_t, \bm{p}_t \rangle - \sum_{t \in I} \langle \bm{g}_t, \bm{p}_t \rangle  \leq \frac{N}{\gamma}\sqrt{T} \sqrt{\frac{1}{2}\ln \frac{1}{\delta'}},
\end{eqnarray}
where we have used the fact that $0\leq \langle \bm{\hat{g}_t}, \bm{p}_t \rangle \leq \frac{N}{\gamma}, \forall t \in I,$ and $|I| \leq T.$ 
Using the same argument, it also follows that with probability at least $1-\delta'$, we have 
\begin{eqnarray} \label{mg-bd2}
	\sum_{t \in I} \langle \bm{g}_t, \bm{p}^\star \rangle - \sum_{t \in I} \langle \bm{\hat{g}}_t, \bm{p}^\star \rangle  \leq \frac{N}{\gamma}\sqrt{T} \sqrt{\frac{1}{2}\ln \frac{1}{\delta'}}.
\end{eqnarray}
Summing up inequalities \eqref{ogd-bd}, \eqref{mg-bd1}, and \eqref{mg-bd2}, we conclude that with probability at least $1-2\delta',$ we have 
\begin{eqnarray} \label{bd2}
	\sum_{t \in I} \langle \bm{g}_t, \bm{p}^\star \rangle - \sum_{t \in I} \langle \bm{g}_t, \bm{p}_t \rangle \leq  \frac{N}{\gamma}\sqrt{2T}\bigg(1+ \sqrt{\ln \frac{1}{\delta'}}\bigg).
\end{eqnarray}

	Since Algorithm \ref{mab} plays the OGD policy with probability $1-\gamma$ on every round independently of everything else, and since the  gain vectors are non-negative, we have almost surely
	\begin{eqnarray} \label{exp-bd}
		 \mathbb{E} [\langle \bm{g}_t, \bm{J}_t \rangle |\mathcal{F}_{t-1}] 
		 &=& 	 \langle \bm{g}_t, \mathbb{E}[\bm{J}_t |\mathcal{F}_{t-1}] \rangle \nonumber\\
		 &\geq & (1-\gamma) \langle \bm{g}_t, \bm{p}_t \rangle \nonumber\\
		 &\geq& \langle \bm{g}_t, \bm{p}_t \rangle  - \gamma, 
		 \end{eqnarray} 
		 where in the last step, we have used the fact that $||\bm{g}_t||_\infty \leq 1, \forall t.$
		 Clearly, the sum $\sum_{t} \langle \bm{g}_t, \bm{J}_t - \mathbb{E}[\bm{J}_t|\mathcal{F}_{t-1}] \rangle $ is also a zero-mean Martingale difference sequence. Hence, using Azuma's inequality once again, we conclude that with probability at least $1-\delta':$ 
		 \begin{eqnarray} \label{hp-bd}
		 	\sum_{t \in I} \mathbb{E} [\langle \bm{g}_t, \bm{J}_t \rangle |\mathcal{F}_{t-1}]  - \sum_{t \in I}\langle \bm{g}_t, \bm{J}_t \rangle \leq \sqrt{\frac{T}{2}\ln \frac{1}{\delta'}}. 
		 \end{eqnarray}
		 Substituting the bound from Eqn.\ \eqref{exp-bd} into Eqn.\ \eqref{hp-bd}, we have with probability at least $1-\delta':$
		 \begin{eqnarray} \label{mg-diff}
		 \sum_{t \in I}	\langle \bm{g}_t, \bm{p}_t \rangle - \sum_{t \in I}\langle \bm{g}_t, \bm{J}_t \rangle \leq \sqrt{\frac{T}{2}\ln \frac{1}{\delta'}} + \gamma T, 
		 \end{eqnarray}
		 where we have used the fact that the length of the sub-interval $I$ is at most $T$. Finally, summing up Eqns.\ \eqref{bd2} and \eqref{mg-diff}, we conclude that with probability at least $1- 3\delta',$ we have  
		 \begin{eqnarray*}
		 	\sum_{t \in I} \langle \bm{g}_t, \bm{p}^\star \rangle - \sum_{t \in I}\langle \bm{g}_t, \bm{J}_t \rangle  \leq  \frac{N}{\gamma}\sqrt{2T} + \frac{3N}{\gamma}\sqrt{T \ln \frac{1}{\delta'}} + \gamma T.
		 \end{eqnarray*}
		 The LHS of the above inequality can be identified with the regret $R_I(\bm{p}^\star)$ corresponding to the fixed sub-interval $I$ and the benchmark $\bm{p}^\star.$ Upon 
		 choosing $\gamma = N^{\nicefrac{1}{2}}T^{-\nicefrac{1}{4}}$ for $T \geq N^2$ yields with probability at least $1-3\delta':$
		\begin{eqnarray*}
			R_I(\bm{p}^\star) \leq 3 \sqrt{N}T^{\nicefrac{3}{4}} \bigg(1+ \sqrt{\ln \frac{1}{\delta'}}\bigg).
		\end{eqnarray*} 
		 Finally, taking a union bound over at most $T^2$ possible sub-intervals $I$ and $N$ possible benchmark actions and redefining $\delta \equiv 3\delta' NT^2,$ we obtain 
		 \begin{eqnarray*}
		 	\max_{I \subseteq [T]} R_I^{\textrm{ALG 2}} \leq 3 \sqrt{N}T^{\nicefrac{3}{4}} \big(1+ \sqrt{\ln \frac{3NT^2}{\delta}}\big),  \textrm{w.p. at least } 1-\delta.
		 \end{eqnarray*}
		 
		 \end{IEEEproof}

\section{Numerical Simulations} \label{sims}
\begin{figure}[!htb]
    \centering
    \includegraphics[width=0.78\linewidth]{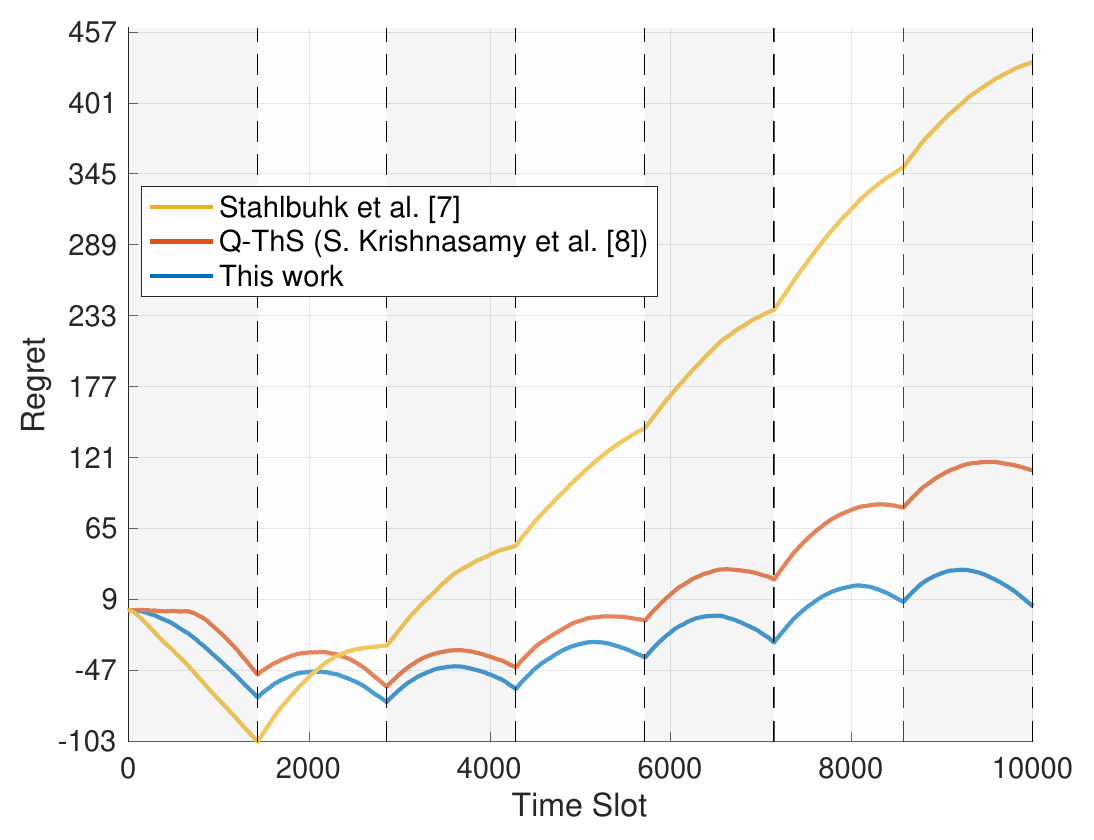}
    \caption{\small{Comparison of queue length regret of Algorithm \ref{mab} with \cite{krishnasamy2016regret}, \cite{stahlbuhk2021learning} in a non-stationary environment where \( T \) is divided into \( m=7 \) blocks, highlighted with alternating backgrounds.}}
    \label{fig:qregret_plot}
\end{figure}

We consider a problem instance with \( N = 5 \) channels and \( T = 10^4 \) slots. Arrivals are sampled from a uniform distribution with rate \( \lambda \). The rate $\lambda$ is set as \( \epsilon = 0.05 \) less than the mean of the channel with the maximum transmission rate. The experiment is repeated for 500 runs, and the average queue length regret is plotted in Figure \ref{fig:qregret_plot}.

\paragraph*{Channel Model} The wireless channels are modelled using a non-stationary Markovian process. The time horizon \( [T] \) is uniformly divided into \( m \) blocks, each with a constant duration. For each block $b(t) \in [m],$ the coefficient \( \alpha_i (b(t)) \sim \textrm{Unif}(0,1) \) is independently initialized at the beginning of the block $b(t)$ and remains fixed throughout the block. The current transmission rate of each channel depends on its previous rate in a Markovian fashion as follows:
\[
S_i(t+1) = \max(0, \min(1, \alpha_i (b(t)) \cdot S_i(t) + \zeta_i(t))).
\]
Here, \( S_i(t) \) is the transmission rate for channel \( i \) at time \( t \) and \( \zeta_i(t) \sim \text{Unif}(-1,1) \) is an i.i.d. additive noise term. The max and min operators ensure that the transmission rates stay within the range \( [0, 1] \). 
Due to the non-stationarity across blocks, the regret-minimizing Q-ThS algorithm from \cite{krishnasamy2016regret} fails to adapt quickly because its performance is heavily dependent on the stationarity of the channel statistics. The same issue arises for the algorithm in \cite{stahlbuhk2021learning}, which has the additional disadvantage that its exploration is restricted to periods when the queue is empty.
In contrast, our algorithm quickly captures the distribution shift in the channel statistics. The simulation code has been made publicly available \cite{qregretcode2025}.

\section{Conclusion} \label{conclusion}
In this paper, we considered an online scheduling problem for a single transmitter-receiver pair to minimize the queue length regret under an adversarial model. We proposed a weakly adaptive MAB algorithm and showed that it yields $\tilde{O}(\sqrt{N}T^{\nicefrac{3}{4}})$ queue length regret w.h.p. In the future, it will be interesting to close the gap between the upper bound obtained in this paper and the lower bound of $O(\sqrt{NT})$.
\section{Appendix} \label{appendix}
\begin{lemma} \label{lindley_proof}
	Let $\{b_t\}_{t \geq 1}$ be sequence of real numbers. Consider the recursion 
	\begin{eqnarray} \label{recursion}
			Q(t)= \big(Q(t-1)+b_t\big)^+, ~t\geq 1, ~~ Q(0)=0.
	\end{eqnarray} 
	Then $Q(t)= \max_{0 \leq t' \leq t} \sum_{\tau=t'}^t b_\tau.$ 
\end{lemma}
\begin{IEEEproof}
	Fix any $1\leq t' \leq t.$ For any $ 1\leq \tau \leq t,$ we have 
	\begin{eqnarray*}
		Q(\tau) \geq Q(\tau-1)+b_\tau. 
	\end{eqnarray*}
Summing up the above inequalities for $ t' \leq \tau \leq t,$ we conclude 
\begin{eqnarray} \label{ineq1}
	Q(t) \geq Q(t'-1) + \sum_{\tau=t'}^t b_\tau \geq  \sum_{\tau=t'}^t b_\tau,
\end{eqnarray}	
where the last inequality follows as $Q(t'-1) \geq 0$. Since Eqn.\ \eqref{ineq1} holds good for any $1\leq t' \leq t,$ we have 
\begin{eqnarray} \label{lb}
	Q(t) \geq \max_{1\leq t' \leq t}  \sum_{\tau=t'}^t b_\tau.
\end{eqnarray} 
To establish the other direction, let $0\leq t^\star \leq t$ be the largest time for which 
$Q(t^\star)=0.$ Note that $t^\star$ is well-defined as $Q(0)=0.$ Since $Q(\tau)>0, t^\star+1 \leq \tau \leq t,$ from Eqn.\ \eqref{recursion}: 
\begin{eqnarray*}
	Q(\tau)=Q(\tau-1)+b_\tau, ~ t^\star+1 \leq \tau \leq t.
\end{eqnarray*}
Summing up the above equations, we conclude that 
\begin{eqnarray} \label{ub}
	Q(t) = \sum_{\tau=t^\star+1}^t b_\tau \leq \max_{1\leq t' \leq t}  \sum_{\tau=t'}^t b_\tau.
\end{eqnarray}
Eqns. \eqref{lb} and \eqref{ub} together conclude the proof. 

\end{IEEEproof}

\clearpage
\bibliographystyle{IEEEtran}
\bibliography{OCO}

\end{document}